# Visualizing incommensurate inter-valley coherent states in rhombohedral trilayer graphene

Yiwen Liu[1#], Ambikesh Gupta[1#], Youngjoon Choi[2#], Yaar Vituri[1], Hari Stoyanov[2], Jiewen Xiao[1], Yanzhen Wang[1], Haibiao Zhou[1], Barun Barick[1], Takashi Taniguchi[3], Kenji Watanabe[4], Binghai Yan[1], Erez Berg[1], Andrea F. Young[2], Haim Beidenkopf[1*], Nurit Avraham[1*]

[1]Department of Condensed Matter Physics, Weizmann Institute of Science, Rehovot 7610001, Israel

[2]Department of Physics, University of California at Santa Barbara, Santa Barbara, CA, USA.

[3]Research Center for Electronic and Optical Materials, National Institute for Materials Science, Tsukuba, Japan.

[4]Research Center for Materials Nanoarchitectonics, National Institute for Materials Science, Tsukuba, Japan.

\# These authors contributed equally

†nurit.avraham@weizmann.ac.il; haim.beidenkopf@weizmann.ac.il

ABC-stacked rhombohedral graphene multilayers exhibit a wide variety of electronic ground states characterized by broken isospin symmetry and superconductivity[1–11]. Recently, indirect evidence of inter-valley coherent (IVC) order has been reported in rhombohedral trilayer graphene (RTG)[3], with possible implications for the origin of superconductivity[12–14]. Here, we report the direct visualization of IVC order in RTG using scanning tunneling microscopy and spectroscopy. Tuning the chemical potential through the Van Hove singularity near the edge of the valence band, we observe a cascade of phase transitions associated with the formation of half- and quarter-metal states. IVC phases, distinguished by an enlarged real space unit cell, are directly imaged near both the high- and low-density boundaries of the half-metal phase. At high hole density, we precisely reconstruct the IVC band structure through quasiparticle interference. Intriguingly, the charge density modulations reveal a $C_3$-symmetric incommensurate IVC order that agrees with the recent prediction of an IVC-crystal phase. Our findings demonstrate that IVC phases are a widespread symmetry-broken ground state within graphene systems.

Rhombohedral graphene multilayers have emerged as a fascinating platform for exploring a wide range of correlated electronic phenomena due to the promotion of strong electron-electron interactions within gate-tunable van Hove singularities (VHSs)[1–11,15–35]. Central to the correlated electron problem in these systems is the spontaneous lifting of the `isospin' degeneracy, which includes the electron spin as well as the valley degree of freedom native to honeycomb systems. Experimentally, changing the electron density $n_e$ and tuning the band structure via the displacement field $D$ leads to the emergence of diverse competing symmetry-broken metallic phases[2], interspersed with superconducting pockets at the lowest temperatures[1]. A number of phases host magnetic order that is directly amenable to macroscopic experimental probes. For example, fully or partially valley imbalanced phases are accompanied by finite orbital magnetization and anomalous Hall responses[2–5,16–18,21–28,32–34]. Other proposed symmetry-broken phases, however, are characterized by a lack of net magnetic moments, instead of being distinguished by lattice scale charge orders.

One such phase is the intervalley coherent (IVC) state, in which the electronic wave functions in the two valleys form a superposition characterized by a macroscopically coherent phase[12–14,36–44]. This phase folds the Brillouin zone by the associated inter-valley scattering wavevector, tripling the real space unit cell size. IVC phases have been previously observed using direct spatial imaging of the tripled-unit cell via scanning tunneling microscopy (STM) in a variety of graphene-based systems, including twisted graphene, multilayer graphene as well as monolayer graphene in the quantum Hall regime[45–52]. In RTG, however, only indirect evidence for the existence of IVC order was reported within a portion of the spin- and valley- polarized quarter metal phase[3]. Theoretically, however, IVC orders of several types have been predicted to exist within the density- and displacement field-tuned phase diagram[12,13,37,38]. Intriguingly, superconducting domes seem to emerge at the boundaries of these predicted phases, possibly suggesting that fluctuations of the IVC order playing a role in electron pairing[12–14].

Here, we provide the first direct observation of IVC states in RTG using STM and scanning tunneling spectroscopy (STS). The IVC phases are observed at both high- and low-hole densities, surrounding the 'half-metal' state residing at the intermediate hole doping. The correlated band structure and IVC states are visualized in both momentum-resolved quasiparticle interference (QPI) patterns, and in real space by direct imaging of surface charge density modulations. In momentum space, the hybridization between the two valleys induces intra-IVC band scatterings, which are evident in the multiple scattering modes reconstructed through Fourier transform (FT) analysis of QPI. By systematically mapping the local wavefunction distribution, we identify intricate charge density wave (CDW) patterns associated with the IVC order and visualize their evolution with energy, carrier density and displacement field. We find the maximum IVC prominence around Fermi energy ($E_\text{F}$) at two separate density regimes in the phase diagram. At high hole doping, our observations support the existence of an incommensurate IVC-crystal state, which has not been observed in other graphene-based systems.

**Gate-tunable spectroscopy of RTG**

The measurement setup is illustrated in Fig. 1a. Our device is fabricated by placing a graphene monolayer on top of a Bernal-stacked graphene bilayer with a small angular misalignment (0.05°) between the mono- and bilayers (see Methods for the fabrication procedures)[31]. This tiny twist angle causes the system to relax into alternating domains of ABA and ABC stacking configurations, separated by domain walls. Since the Bernal (ABA) stacking represents the lowest energy stacking configuration[53–55], the system relaxes into shrunken triangular ABC domains surrounded by larger regions of ABA stacking (Fig. 1a). Remarkably, we find that the energy landscape between the two stacking configurations can be tuned by the back gate voltages ($V_{bg}$) (see Method and Supplementary Note 1). The free-energy difference between the two stacking orders scales quadratically with the electric field, allowing the ABC stacking to become more favorable with increasing the displacement field[56,57]. This enabled us to manipulate and expand the ABC stacking area controllably, and eventually stabilize it by increasing $V_{bg}$[57,58]. The topographic features of the ABC region are shown in Fig. 1b. A model side view and an atomically resolved topographic image are shown at the bottom of Fig. 1a. In this staggered rhombohedral arrangement, the $A_1$ atoms on the top layer are above vacant sites in the middle layer, experiencing lower interlayer coupling than the $B_1$ atoms located directly above $A_2$ atoms, which causes the wave function near zero energy to be predominantly localized on the $A_1$ and $B_3$ lattice sites of the two outer layers. As a result, primarily the $A_1$ lattice can be visualized in low-energy surface state-sensitive STM measurements (bottom left panel of Fig. 1a).

Further identification of the local stacking order is obtained by comparing the spectra of the two domain regions. Figure 1c shows spectra of ABC and ABA stacking configurations measured at a high displacement field in the triangular region and its surroundings. While the ABA spectrum remains semi-metallic, displaying a subtle peak at the band edge (blue arrow), the ABC spectrum exhibits a distinct energy gap, accompanied by an ultra-sharp Van Hove singularity (VHS) peak (see Fig. 1d for the corresponding band structure), indicative of the hole-side flat band below the $E_F$ (Supplementary Note 2). In particular, the application of a displacement field influences both the band structure and the wave function distribution across the layers. This tunability is demonstrated in Fig. 1e through local density of states (LDOS) measurements while varying bottom gate voltages. In this $dI/dV$ ($V_{bg}, V_{bias}$) map, two phenomena are observed. First, the energy difference between VHSs of the conduction and valence bands enlarges with the increase of displacement fields. Second, the VHS of the valence band becomes more pronounced at negative gate voltages while the VHS of the conduction band intensifies at positive gate voltage, exhibiting a polarity change near zero $V_{bg}$.

To understand the features of this $dI/dV$ ($n_e, D$) map, we convert the gate voltage $V_{bg}$ into carrier density $n_e$ and displacement field $D$[59–61], where the back-gate to RTG capacitance is deduced from Landau level (LL) spectroscopy (see Methods and Fig. S3). Zero displacement field is hallmarked by the minimal energy separation between the VHSs of the conduction and valence bands which occurs at zero $V_{bg}$ (Fig. S3), suggesting that there is a negligible tip-induced electric field without applying back-gate voltage. In contrast, we find a finite electron density at zero back-gate ($n_0$), and the mid-VHSs energy occurs at zero bias ($E_F$) under a finite negative gate voltage -2.8 V, which we set as the charge neutrality point (CNP). This enables us to compare the $dI/dV$ map to the top layer DOS via tight-binding (TB) calculations (Fig. 1f). All the above features are remarkably reproduced (see Methods for the details of the fittings). To further understand the polarity change, we project the calculated wavefunction of the $A_1$ sublattice in the top layer onto the band structure (Fig. 1g) for negative, zero, and positive displacement fields. At $D = 0$, the low energy states of the conduction and valence bands have approximately equal weight on the top surface (middle panel). However, when a finite displacement field is applied, the inversion symmetry of the system is broken and the wavefunction distribution changes: a negative $D$ causes the low energy states of the valence band to be predominantly localized on the top surface (bottom panel) and vice versa for positive $D$ (top panel). Indeed, we find that at zero $V_{bg}$, the valence- and conduction-VHS peaks have similar intensities (Fig. 1e), suggesting a zero net interlayer potential and marking the switch point of the wavefunction distribution across the layers (middle panel in Fig. 1g).

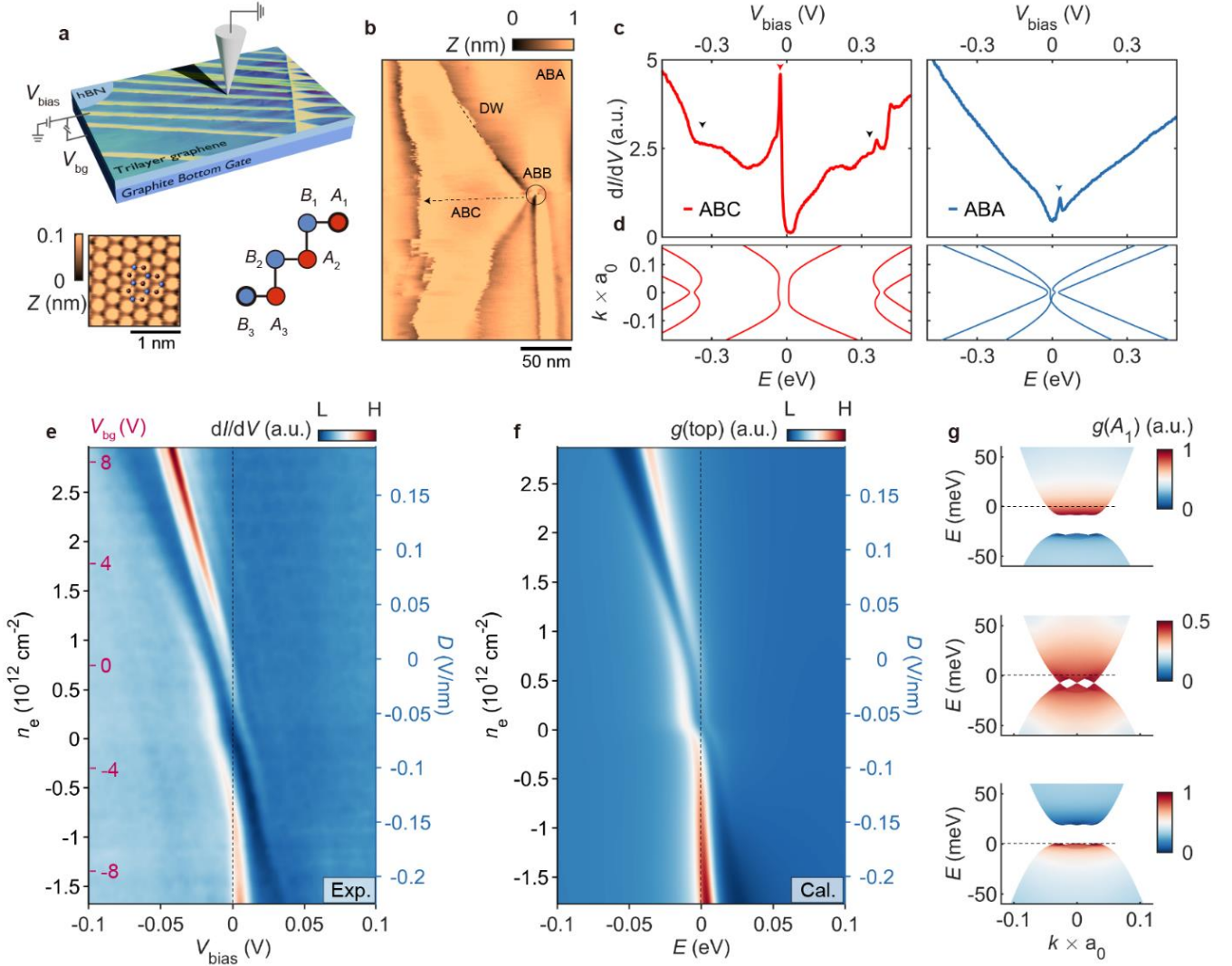

**Fig. 1 | Experimental setup and gate-tunable RTG spectroscopy. a,** Configuration of the STM measurement. Trilayer graphene is supported by a bottom graphite gate for controlling both carrier density $n_e$ and displacement field $D$. The back gate $V_{bg}$ is added on graphite and the $V_{bias}$ is applied onto the sample. A conductive AFM (cAFM) image, obtained by a different setup, is placed on top of the hBN layer to show the large-scale features of the trilayer graphene. The real space lattice structure obtained by STM and the RTG lattice configuration are shown in the bottom left and right insets. The atoms in each layer are offset relative to the layer below by one third of the graphene unit cell, resulting in the inversion symmetric structure across the layers. **b,** STM topographic image of A-AB RTG with a typical size of around 150 nm by 350 nm in the minor and major directions. **c,** STS of ABA and ABC stacking domains obtained at $T = 4$ K and high displacement field. At higher energies (±350 mV), we observe a step-like feature in ABC-stacked region (black arrows), which corresponds to the remote bands in the RTG band structure. **d,** Calculated band structures of ABC and ABA stacking orders with interlayer potential $\Delta_1 = 20$ mV. Here $a_0 = 0.246$ nm is the graphene lattice constant. **e,** d$I$/d$V$ spectra as a function of bias voltage $V_{bias}$ and bottom gate voltage $V_{bg}$ (marked by red labels), measured at the center of ABC-stacked domain at $T = 4$ K. **f,** Calculated DOS for the top graphene layer as a function of $D$, $n_e$ and energy $E$. **g,** Projected DOS $g(A_1)$ with interlayer potential $\Delta_1 = \pm 10$ meV (top and bottom), 0 meV (middle) onto the $A_1$ lattice of the top layer. The bands are adjusted to align with the simulation in panel f.

**Momentum space visualization of an IVC state**

Signatures of electronic correlations are primarily anticipated when the Fermi level (indicated by $V_{\text{bias}} = 0$) approaches the flat band. Such correlations would manifest as deviations from the single-particle band structure. As shown in Fig. 1e, at $V_{\text{bg}}$ = 0 V, RTG is slightly electron-doped, with $E_F$ positioned just above the VHS of the conduction band, where the LDOS is relatively low, hence no correlation effects are expected. In contrast, we find an extended Fermi level pinning of the hole-doped VHS between $V_{\text{bg}}$ = -3 V and -9.5 V (the maximal gate voltage we can apply). To explore electronic correlation effects, we measure the energy dispersion of the band structure at $V_{\text{bg}}$ = 0 V and -9.5 V, using quasiparticle interference (QPI) imaging[62]. In this technique, quasiparticle scattering from atomic step edges or impurities leads to periodic interference patterns in real space, whose FT can be related to elastic scattering processes within the band structure in reciprocal space[63,64]. Such interference patterns emanating from the ABB-stacked node are acquired along the line in Fig. 1b, at $V_{\text{bg}}$ = 0 V and -9.5 V, are shown in the $dI/dV$ maps (Figs. 2a and 2b). Notably, we observe a distinct dispersing interference pattern away from the point scatterer, superimposed on non-dispersing LDOS modulations in both the conduction and valence bands (see Methods and Supplementary Note 4).

The FTs of Figs. 2a and 2b are shown in the left panels of Figs. 2c and 2d, respectively. They display the energy dispersion of the scattering wave vectors alongside the single-particle joint density of states (JDOS) in the right panels, calculated by autocorrelating the TB band structure. Here, the interlayer potential for each gate voltage is adjusted to match the band edges, marked by the nondispersive horizontal line-patterns (red dashed lines) in the momentum-resolved QPI. Notably, in both gate voltages the main dispersive scattering branches of the conduction and valence bands, along with the increasing band gap with $D$ matches the single-particle JDOS calculations (Figs. 2c and 2d right panels). However, a closer examination of the scattering branches reveals a significant distinction: at $V_{\text{bg}} = 0$ V, we observe a single scattering branch for the valence band, while at $V_{\text{bg}} = -9.5$ V, where the valence band VHS is pinned at $E_F$, we clearly resolve multiple scattering branches (red arrows in Fig. 2d, left panel), contrasting with the JDOS calculation which predicts only one mode (Fig. 2d right panel). This splitting of the scattering branches is highlighted in Fig. 2e (left), zooming in on the relevant energies of the valence band scattering enclosed in the rectangle of Fig. 2d. Two distinct QPI modes (red and black arrows) are clearly visible along with a fainter third mode (blue arrow). The existence of these modes is further corroborated by the phase signal of the FT (Fig. 2e, right). The third mode is less distinct than the other two, and we capture a glimpse of it among other peaks in FT at the vicinity of the VHS at zero bias.

The emergence of multiple QPI modes hallmarks a departure from the single-particle picture due to the onset of strong electronic interactions. When the Fermi level is tuned within the flat band, the degenerate states become less favorable, driving the system into a symmetry-broken state and creating additional scattering channels. Notably, the extracted energy dispersion of the three modes (Fig. 2f) shows an interesting trend: the momentum separation between the modes increases as $|E|$ increases. This behavior is captured in our self-consistent Hartree-Fock (SCHF) band structure calculation (see Methods)[13], which predicts an IVC phase as a ground state for the corresponding hole density. As depicted in Fig. 2g, the IVC band structure allows for three distinct scattering processes, marked by red and blue (**q₁** and **q₃**, intervalley scattering) and black (**q₂**, intravalley scattering) arrows. The energy dispersion of the three calculated scattering channels (solid lines in Fig. 2f) closely matches that of the measured channels. We note that the simple isospin splitting without IVC would predict a different energy dispersion trend (see Supplementary Note 4), supporting our identification of an IVC state in momentum space.

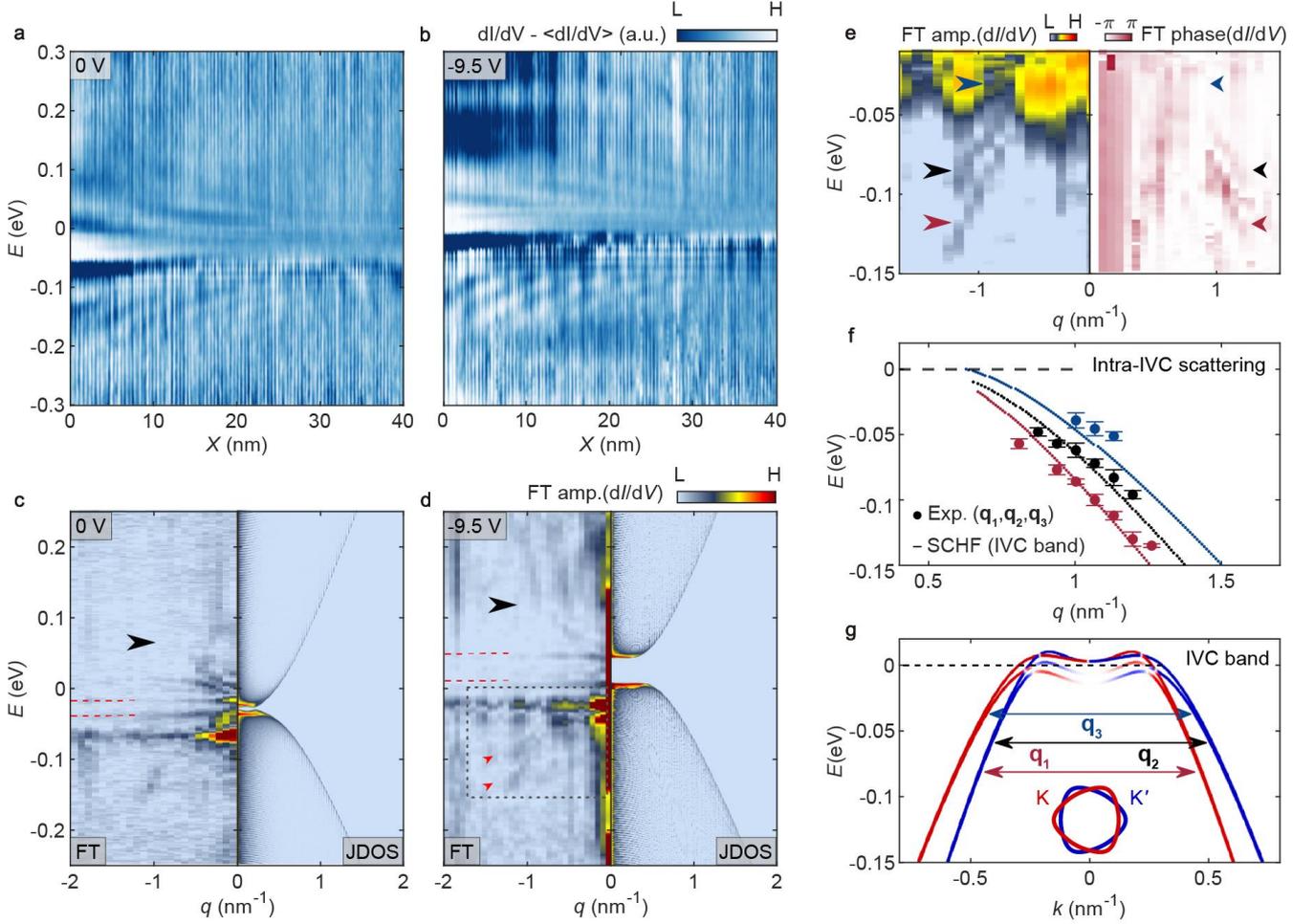

**Fig. 2 | Reconstructed RTG band structure and IVC order in QPI. a,b,** $dI/dV(X, V_{bias})$ maps taken along the dashed line in Fig. 1b, measured at $T$ = 4 K. The periodic vertical lines represent the atomic-scale LDOS variation. **c,** Left: FT amplitude of the $dI/dV$ map in panel a, for the full dashed line in Fig. 1b. The red dashed lines indicate the edges of the conduction and valence bands. Right: JDOS for RTG calculated using a TB model (see Methods for parameters) with an interlayer potential $\Delta_1 = 2$ mV. **d,** Left: FT of the $dI/dV$ map in panel b for the full dashed line in Fig. 1b. Right: Similar to panel c (right) but with $\Delta_1 = 18$ meV. **e,** Left panel: FT amplitude of panel b after normalizing the $dI/dV$ by $I/V$ to emphasize the three dispersing branches, specifically zoomed into the energy range of the valence band. Right panel: The phase counterpart of panel e (left) showing two clear dispersing branches and a slight hint of the third. **f,** Measured energy dispersion of the three scattering processes, extracted from panel e (symbols) along with the calculated SCHF energy dispersion (solid lines) of the three scattering processes shown in panel g. **g,** Band structure of IVC phase calculated via SCHF approximation (K and K' valleys represented by blue and red lines). The arrows indicate permissible scattering processes. Lower inset shows the schematic image of the nesting of the Fermi surfaces centered at K and K' valleys in the IVC phase.

### Real space visualization of IVC orders

In addition to modifying the single-particle band structure, IVC phase can lead to pronounced variations in the LDOS in the form of CDW or Kekulé-type charge orders, where charge density modulation breaks the translational symmetry of the underlying lattice[45–52]. To further investigate the existence of IVC phases and other symmetry-broken states, we performed high-resolution spectroscopic measurements of the LDOS and real space imaging of the local density modulation. In Fig. 3a, we focus on the regime where the valence band VHS is pinned at $E_F$ for a range of carrier densities. Remarkably, we observe a cascade-like pattern in $dI/dV$ spectra, characterized by three distinct splittings occurring around gate voltages of -6 V, -7.7 V, and -9.5 V. Notably, the last one, at -9.5 V, aligns with the IVC splitting identified in our QPI measurements. This cascade pattern is reproducible across different setpoints and cooling cycles (Fig. S5), with consistent results obtained at 4 K and 0.4 K. Similar spectroscopic features have been reported in other graphene based-systems and have been interpreted as a signature of electronic correlations in the form of symmetry breaking[65–75].

Three representative $dI/dV$ spectra (Fig. 3c), were measured at corresponding carrier densities marked by the arrows in Fig. 3a. The spectra reveal three distinct energy splittings of the VHS with varying magnitude: 2.3 mV, at low hole density (top), where the peaks surround $E_F$ and a third one appears 1 meV below, 1.8 mV at intermediate hole density where the peaks surround $E_F$ (middle), and 2.5 mV at high hole density where only the lower peak overlaps with $E_F$ (bottom). The 1.8 mV splitting at $V_{bg}$ = -7.7 V reveals nearly equivalent $dI/dV$ peak intensity below and above the Fermi level, indicating that two out of the four flavors are occupied (see Fig. S9 for other regions)[42]. Tracking the zero bias $dI/dV$ as a function of $V_{bg}$ (Fig. 3b) shows a series of suppressions of the LDOS at $E_F$ around $V_{bg}$ = -6 V, -7.7 V and -9.5 V (marked by the dotted lines). Moreover, the LDOS at $E_F$ never reaches zero, implying that the emergent symmetry-broken phases remain metallic, consistent with recent measurements on RTG and theoretical prediction[2,13,37].

To further characterize the observed splitting and the emergent symmetry broken phases, we examine their associated charge ordering at the atomic scale. Real-space $dI/dV$ maps are presented in Figs. 3d-g alongside their FTs (left and right panels) across a range of $V_{bg}$, where the cascade pattern was detected in STS. For all gate voltages, the triangular graphene $A$ sublattice (Fig. 1a) is clearly visible in the real-space LDOS, accompanied by the corresponding graphene Bragg peaks in the FTs (red arrow). However, at $V_{bg}$ = -6 V, -9 V, and -9.5 V, additional FT peaks (black arrows) emerge within the graphene Bragg peaks. The corresponding charge modulation in real space has a wavelength of 4.3±0.2 Å (see Fig. S6), consistent with an approximate tripling of the unit cell due to the IVC ordering. At $V_{bg}$ = -6 V (Fig. 3d), the IVC modulation is relatively weak and only detectable in the FT image (black arrow). At $V_{bg}$ = -9 V and -9.5 V (Figs. 3f and 3g), however, the modulation is more pronounced and can be resolved in both real and momentum space. Notably, no IVC ordering is detected at $V_{bg}$ = -8 V (Fig. 3e) in between these IVC phases, even though the cascade is present, and isospin symmetry is broken.

The phase diagram constructed from our spectroscopic and real-space imaging is as follows: increasing the hole density, the system first enters an IVC phase (IVC$_1$), characterized by the splitting of VHS into multiple peaks and a weak charge density modulation. This is followed by an isospin symmetry-broken phase with equal intensity peaks above and below $E_F$ and no observable real-space modulation. Finally, at higher hole doping, the system transitions into another IVC phase (IVC$_2$), exhibiting a stronger CDW. Previous capacitance and magnetometry report a transition from an IVC quarter metal to a spin-polarized half-metal to a partially isospin polarized phase as the hole density increases[2,3], in line with our identification. We note, however, that our density-displacement field trajectory (Fig. S10) does not exactly cross the same phase spaces as in prior results. This discrepancy may be attributed to different dielectric screening strengths in dual-gated and single-gated open-surface device geometry[76].

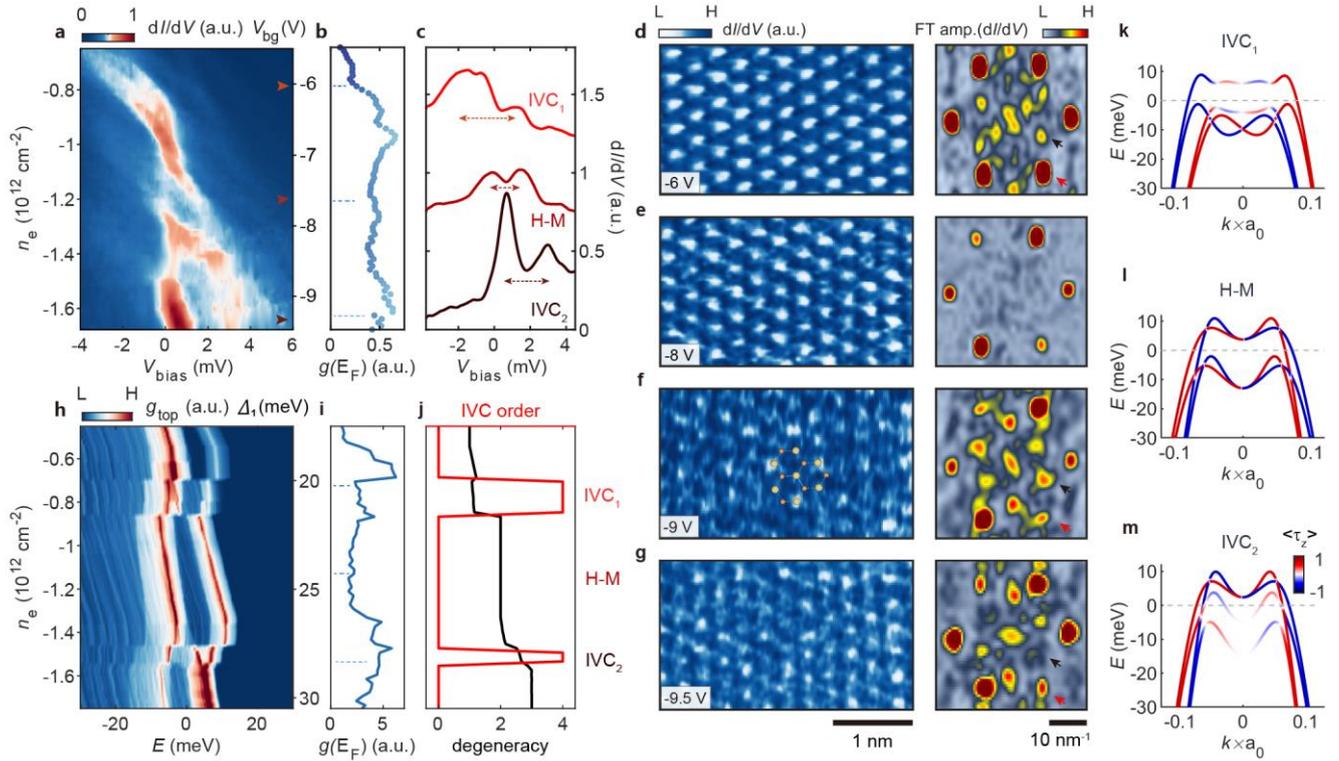

**Fig. 3 | Spectroscopic visualization, modelling of symmetry breaking and mapping of IVC orders in real space. a,** Spectra acquired at the center of the ABC-stacked domain, $T = 0.4$ K. Triangles mark the corresponding gate voltages for the STS in panel c. **b,** Gate-dependent LDOS at $E_F$ extracted from panel a. The dashed lines mark $V_{bg}$ for the corresponding STS in panels a and c **c,** Three representative $dI/dV$ curves from panel a for the corresponding gate voltages: $V_{bg}$ = -6 V, -7.7 V, -9.5 V (top to bottom). The curves are offset for clarity. **d-g,** Left panels: Real space $dI/dV$ maps taken at various $V_{bg}$ and $V_{bias}$ at 4 K. From top to bottom: $V_{bg}$ = -6 V ($V_{bias}$ = -2 mV), -8 V ($V_{bias}$ = 0.1 mV), -9 V ($V_{bias}$ = 1.4 mV), -9.5 V ($V_{bias}$ = 1.2 mV). Right panels: FTs of the left $dI/dV$ maps. **h** Calculation of top layer DOS across a range of densities and displacement fields, extracted from SCHF. **i,** DOS at $E_F$ extracted from panel h. **j,** IVC order and degeneracy as a function of the carrier density. **k-m,** Quasiparticle band structures for hole doping at $n_e$ = -0.79 × $10^{12}$ cm$^{-2}$ (IVC$_1$), -1.13 × $10^{12}$ cm$^{-2}$ half-metal (H-M) and $n_e$ = -1.51 × $10^{12}$ cm$^{-2}$ (IVC$_2$). Line color corresponds to the valley polarization of the quasiparticle. A varying $\langle \tau_z \rangle$ between -1 and 1 corresponds to an IVC band.

To theoretically understand the emergence of the cascade of phase transitions and IVC orders, we project the DOS obtained through SCHF calculations onto the top layer, along a diagonal trajectory within the phase diagram, as shown in Fig. 3h (see schematic phase diagram in Fig. S10). Increasing the hole density from CNP induces successive phase transitions in RTG. The system first enters a quarter-metal phase around $n_e$ = -0.6 × $10^{12}$ cm$^{-2}$, then a half-metal phase around $n_e$ = -1.2 × $10^{12}$ cm$^{-2}$, and eventually a three-quarter-metal phase at the high hole doping, by occupying 1, 2, and 3 of the isospin flavors, respectively. The detailed degeneracy of each phase is plotted in black on Fig. 3j. For the half-metal phase, the DOS shows equal intensity for both occupied and unoccupied states, which qualitatively agrees with our experimental results shown in Fig. 3c.

Significantly, distinct IVC phases are interspersed throughout the phase transition sequence (see the extended DOS in Fig. S10), as evidenced by the splitting into multiple VHSs around the $E_F$ (Fig. 3h). Beyond the band-splitting on either side of the Fermi level, IVC$_1$ shows two peaks below $E_F$, while IVC$_2$ displays two more pronounced peaks above $E_\mathrm{F}$, which qualitatively aligns with the experimental result, showing multiple splittings around $E_\mathrm{F}$ (Fig. 3c). To better illustrate the symmetry-broken phases at different densities, the calculated band structures are shown in Figs. 3k-m. In the half-metal phase (Fig. 3l), there are spin-polarized and valley-unpolarized splitted bands, with two-hole pockets above $E_\mathrm{F}$. In STM measurements, the empty states above $E_\mathrm{F}$ can also be probed, allowing us to resolve the two VHSs corresponding to spin-up and spin-down or K/K' states (Fig. 3c). For the IVC order, hybridization of the two valleys via the density-density interaction causes a band splitting around $E_\mathrm{F}$, as illustrated in Figs. 3k and 3m. Furthermore, the calculated top-layer DOS at $E_\mathrm{F}$ (Fig. 3i) aligns with the measured zero bias LDOS intensity presented in Fig. 3b. This provides additional confirmation of the relative positioning within the phase diagram observed in our experiments, involving a transition from the IVC phase to a half-metal phase and then back to the IVC phase (Fig. 3j).

**Incommensurate IVC-crystal phase**

To pinpoint the exact periodicity of the IVC Bragg peaks and assess their commensurability with the graphene structure, we present in Fig. 4a (also see Fig. S7) FT images with higher momentum resolution taken at $V_\mathrm{bg}$ = -9.5 V for two $V_\mathrm{bias}$, where the average intensity of the IVC Bragg peaks is strong (Fig. 4b). Here, the Bragg peaks of both the graphene and the IVC lattices appear significantly sharper, enabling precise determination of their wavevectors. The positions of the graphene peaks and the IVC Bragg peaks are marked by the outer blue and the inner red circles, while the inner-most gray circle marks the expected commensurate IVC Bragg peak positions at a radius of $1/\sqrt{3}$ times that of the outer graphene circle. Notably, the wavevectors of all the measured IVC Bragg peaks have magnitudes that exceed those expected from a commensurate IVC. To quantify the level of incommensurability, we plot in Fig. 4c the angularly averaged FT intensity of Fig. 4a (left panel). The average graphene Bragg peak is located at $\mathbf{G}_\mathrm{G}$=25.5±1.5 nm⁻¹ (the error range represents half width at half max) setting the commensurate IVC Bragg peaks, as derived from the graphene peaks (orange curve), at $\mathbf{G}_\mathrm{c}=\mathbf{G}_\mathrm{G}/\sqrt{3}$=14.7±0.9 nm⁻¹. However, the IVC Bragg peak we find in our experiment appears at $\mathbf{G}_\mathrm{IVC}$=18.5±1.0 nm⁻¹, signifying the incommensurate nature of the IVC state we image at high hole density with a characteristic wavevector offset of $q=G_\mathrm{IVC}-G_\mathrm{c}$~3.8±1.3 nm⁻¹. Despite some overlap between the two peaks (IVC and $\mathbf{G}_\mathrm{G}/\sqrt{3}$), the incommensurability remains unambiguous. This large wavelength super-modulation of the IVC pattern can explain the rather complex structure we find in the real space charge distribution in Fig. 3f. We stress that the incommensurate IVC state remains fairly $C_3$-symmetric with all peaks shifted outwards from the commensurate momenta.

Recent theoretical studies indeed predict a stable $C_3$-symmetric incommensurate IVC-crystal phase[13]. By optimizing quasi-nesting within the annular Fermi surface, the scattering wavevector, $\mathbf{K}-\mathbf{K'}+\mathbf{q}$, is found to connect the outer corner and the inner edge of the warped annulus that have opposite Fermi velocities[77], as depicted in Fig. 4d. The folded Fermi surface hybridizes all K' valleys with K symmetrically, as shown in the inset of Fig. 4e. To the best of our knowledge, this is the only existing model that accommodates a $C_3$-symmetric incommensurate IVC state in RTG. In general, the IVC-crystal is expected to give rise to three incommensurate IVC Bragg peaks at momenta $\mathbf{q}_i$ arranged symmetrically around $\mathbf{G}_\mathrm{c}$, as shown in Fig. 4e. These include an outer peak that results from direct scattering between the inner and outer edges of the annular Fermi surface (red arrow), and two inner ones which result from secondary scattering by a reciprocal lattice wavevector $\mathbf{G}$ (blue arrow). However, our calculations reveal that the relative intensities of the three predicted peaks should not be equal, and in fact are determined by the wavefunction structure at a small momentum $\mathbf{q}$ away from the K valley. The deviation from equal intensity at the three wavevectors is predicted to be of the order of the relative wavefunction weight on the $A_1$ and $B_1$ lattice sites (Supplementary Note 10). In monolayer graphene, the ratio $\psi_{1,B_1}/\psi_{1,A_1}$ is approximately 1, allowing for a significant modulation in the relative intensity. In contrast, in RTG the $B_1$ atom is tightly bonded to the $A_2$ atom, hence the ratio obtained by TB models is of the order of $\psi_{1,B_1}/\psi_{1,A_1} \sim 0.05$, which makes it difficult to explain why only one of the three peaks is observed.

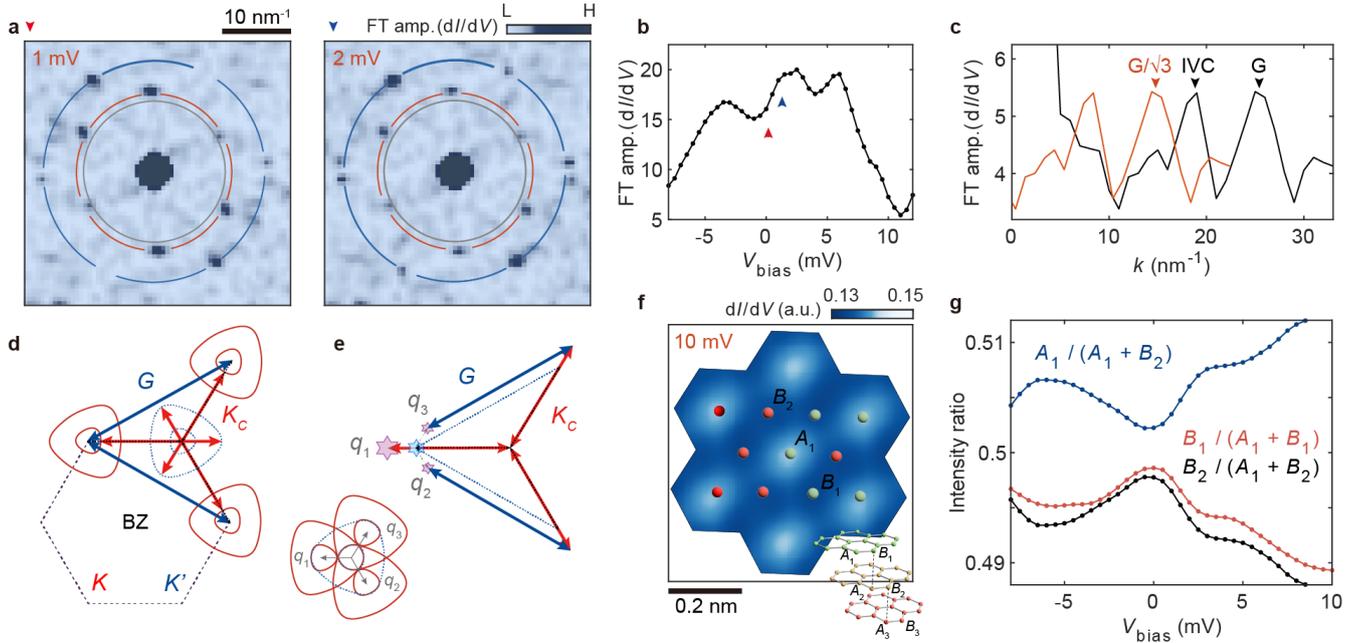

**Fig. 4 | Incommensurate IVC-crystal state at high hole density. a,** The FTs of the larger area map (approximately 7 × 7 nm), providing higher momentum resolution at two low biases, measured at $T$ = 0.4 K. The outer blue circle indicates the position of the graphene Bragg peaks, while the red circle highlights the locations of the IVC Bragg peaks. Additionally, the inner grey circle marks the commensurate IVC momentum calculated from the graphene peak. The FT of the $V_{bias}$ = 2 mV image shows stronger graphene Bragg peaks than the 1 mV FT image. **b,** Bias voltage dependence of the average IVC Bragg peak intensity in the FT images. The intensity of each Bragg peak is calculated by summing the values within a radius of 1 pixel around the graphene or IVC Bragg peaks. **c,** Angular averaged FT intensity of the left panel in a (black line) revealing the locations of both the IVC Bragg peak at $G_{IVC}$=18.5±1.0 nm$^{-1}$ and the graphene Bragg peak at $G_G$=25.5±1.5 nm$^{-1}$. The orange line is generated by normalizing the momentum of the former curve by dividing it by $\sqrt{3}$, finding the expected position of the commensurate IVC Bragg peaks at 14.7±0.9 nm$^{-1}$, with the peak tail overlapping the experimentally observed IVC Bragg peaks. **d**, Annular pockets at K and K' valleys (solid red and dotted blue, respectively). The sizes are exaggerated for clarity. The incommensurate scattering wavevectors and reciprocal wavevectors are shown by red and blue arrows, respectively. **e,** The resulting incommensurate IVC Bragg peaks displaced by **q$_1$, q$_2$, q$_3$** away from the commensurate position (marked by the blue star). The inset shows the incommensurately folded K and K' pockets. **f,** Super-resolved unit cell LDOS distribution at $V_{bias}$ = 10 mV. The graphene lattice is overlayed on top. **g,** Bias dependent intensity of the LDOS at the vicinity of $A_1$, $B_1$, and $B_2$ (blue, orange and black lines). The integration radius of the intensity is 0.05 nm.

Experimentally, however, STM enables us to measure the relative local intensity of the electronic wavefunction on the $A_1$ and $B_1$ sites with high precision, eliminating the need to rely on TB models. To obtain an ultimate spatial resolution, we construct a super-resolved unit cell from our $dI/dV$ maps with the same algorithm used previously in the study of $Bi_2Sr_2CaCu_2O_8$ and more recently of $Co_3Sn_2S_2$[78–80]. We note that due to the incommensurability of the IVC structure and the graphene lattice, it is not possible to construct a super-resolved IVC unit cell together with graphene unit cell. However, our focus here is primarily on the wavefunction distribution within the parent unit cell. A representative LDOS distribution within a super-resolved unit cell is shown in Fig. 4f. It immediately becomes evident that a finite LDOS can be found throughout the unit cell. We thus extract in Fig. 4g its intensity at the immediate vicinity of the $A_1$ and $B_1$ sites across various bias voltages. We find that at low biases, where the IVC-crystal forms, the wavefunction distribution is almost uniform with $\psi_{1,B_1}/\psi_{1,A_1}$ approximately 0.9. This value is closer to the value expected for monolayer graphene than to the small ratio obtained in common TB calculations of RTG. The origin of the discrepancy is yet unknown and may be related to the strong interactions affecting charge localization or even the IVC structure itself. Nevertheless, it supports our observation of suppressed inner IVC-crystal peaks. We thus identify the spontaneously symmetry-broken state at high hole doping as IVC-crystal.

**Discussion**

We report the observation of IVC states at two different regions of the RTG displacement-density phase diagram - at low- and high-hole densities and displacement fields. Between the two regions, the spectrum also displays splittings associated with symmetry breaking, but no translational symmetry breaking is observed, consistent with ferromagnetic states with lifted spin and/or valley degeneracy that maintain conservation of the two valley occupations. The resulting phase trajectory is qualitatively similar to the one measured in dual-gated devices. However, the open-faced structure significantly influences screening, making direct comparisons between the two types of devices difficult. Indeed, for the same values of density and displacement field used in our experiment, the phase diagram of dual-gated devices shows no broken symmetry phases. This can be explained by the potentially stronger effective interaction in the open-faced device studied here, and hence the symmetry-broken phases occupy a larger part of the phase space. An interesting open question is whether superconductivity exists in open-faced devices.

At high hole density, we identified an incommensurate CDW that agrees with an IVC-crystal state[13]. One of the potentially important aspects of this observation is that it generates a mini-Brillouin zone defined by $q_1$, $q_2$, $q_3$ that corresponds to a long-wavelength modulation of the charge density. The mini-Brillouin zone is a salient feature of twisted moiré systems, where it is thought to play an important role in facilitating long-wavelength Umklapp electron-electron and electron-phonon scattering processes that may be important for superconductivity[81–83]. Unlike twisted systems where the moiré unit cell and the resulting mini-Brillouin zone are externally imposed by the atomic structure, in RTG, they both emerge from the spontaneous incommensurate IVC crystal phase we report. We speculate that such an emergent long-wavelength modulation may play a role in enabling superconductivity, either through fluctuations of the IVC order parameter[12,13], or by allowing Umklapp electron-phonon coupling that enhances superconducting pairing[81–83].

Our study raises several questions which remain open for future experimental and theoretical investigation. In our experiment, the energy scales of the observed splittings are significantly smaller compared to the SCHF calculations, which is known to overestimate interaction gaps. However, the discrepancy we find here (by about an order of magnitude) is surprisingly large, calling for calculations beyond Hartree-Fock. Intriguingly, in twisted graphene, the IVC gaps are much larger than our results[46,45]. This difference between the two systems is also reflected in the temperature dependence of compressibility. While the compressibility modulations in twisted bilayer graphene disappear at around 100 K, in contrast, in RTG they persist up to approximately 30 K[9,84]. Another surprising result that warrants further elucidation is the rather uniform distribution of the wavefunction within the RTG unit cell, which is consistent with our observation of the subset of Bragg peaks of an IVC-crystal state.

## Methods

### Sample fabrication and growth of RTG

Given that STM/STS can provide atomic-scale spatial resolution and is immune to global structural disorder, we design RTG by twisting. Trilayer graphene samples were prepared using the dry transfer technique[45]. A single crystalline graphene that consists of monolayer and bilayer is exfoliated on Si/SiO$_2$ substrate. The boundary between the monolayer and the bilayer is then precisely cut using a sharp metallic tip. A few layers graphite for gate is mechanically exfoliated on polydimethylsiloxane (PDMS), then dropped down at 80 ℃ on a prepared hBN exfoliated on Si/SiO$_2$ substrate. Poly (bisphenol A carbonate) (PC) film on a PDMS block is used to pick up the graphite/hBN pair at 120 ℃. With the bottom surface of hBN, the bilayer graphene is picked up first and then the remaining monolayer part is manually rotated by 0.05° and picked up at 40 ℃. To fabricate a STM compatible sample with high surface quality, we used gold-coated PDMS-assisted flipping technique and gold stamping technique for flipping and electrically contacting the prepared stack[45]. Conductive AFM (cAFM, Asylum Research Cypher S with ASYELEC-01-R2) is utilized to identify suitable areas before cooling down the sample. Two regions are imaged using cAFM, highlighted by the squares in Fig. S1.

In addition to twisting, RTG can also be achieved by scanning under high electric fields. Two distinct types of ABC-stacked domains were identified in our experiment: a stable (Fig. 1) and the mobile domain (Figs. S1e-g). For the mobile case, the STM tip is able to tune the ABC domain size, which results from the interplay of pinning forces, elastic energy, stacking energy, and tip-induced effects[57]. By activating the back gate, the ABC domain expanded and stabilized, enabling the acquisition of a clear $dI/dV$ map of the equilibrium configuration. The $dI/dV$ map recorded at the VHS of the valence band after stabilization reveals the homogeneous characteristics of RTG within the triangular region. The corresponding dynamic processes are detailed in Supplementary Note 1.

### STM/STS measurements

STM and STS measurements were performed in UNISOKU (USM-1300) system in an ultrahigh vacuum (UHV) chamber with a pressure of about 10$^{-11}$ Torr. The topographic images were taken under constant current mode by turning on the feedback circuit by adding DC sample bias and back gate. The $dI/dV$ spectra were taken with the standard lock-in technique by turning off the feedback circuit and adding a 793 Hz, the root mean square (RMS) voltage, $V_{\text{rms}}$ modulation through a SR830 lock-in amplifier with the time constant (*TC*) of 3-100 ms depending on the temperature and the required energy resolution, i.e., 4 K, 1 meV; 2 K, 0.6 meV; 0.4 K, 0.1 meV. The voltage divider (1/10 for DC 1/100 for AC modulation) is added when examining the band splitting. The electronic temperature of 0.4 K is 1.1 K in our system as determined by fitting BCS superconducting gap in Indium. The gold electrode was evaporated as the final step in substrate preparation before putting flakes without polymer on top. The measurements were performed with a Pt/Ir tip prepared and characterized on the clean gold electrode. We locate the RTG sample via the capacitance guiding technique reported previously[85].

### Band structure and DOS calculations

We use TB model to obtain the band structure of ABC-stacked trilayer graphene[86]:

$$H_0(k_x, k_y) = \begin{pmatrix} \Delta_1 + \Delta_2 + \delta & v_0\pi^\dagger & v_4\pi^\dagger & v_3\pi & 0 & \frac{1}{2}\gamma_2 \\ v_0\pi & \Delta_1 + \Delta_2 & \gamma_1 & v_4\pi^\dagger & 0 & 0 \\ v_4\pi & \gamma_1 & -2\Delta_2 & v_0\pi^\dagger & v_4\pi^\dagger & v_3\pi \\ v_3\pi^\dagger & v_4\pi & v_0\pi & -2\Delta_2 & \gamma_1 & v_4\pi^\dagger \\ 0 & 0 & v_4\pi & \gamma_1 & \Delta_2 - \Delta_1 & v_0\pi \\ \frac{1}{2}\gamma_2 & 0 & v_3\pi^\dagger & v_4\pi & v_0\pi^\dagger & \Delta_2 - \Delta_1 + \delta \end{pmatrix}, \quad (1)$$

where $\pi = \xi k_x + i k_y$ ($\xi = \pm 1$ corresponds to K and K' valleys, respectively). The Hamiltonian is written in the basis $(A_1, B_1, A_2, B_2, A_3, B_3)$, where $A_i/B_i$ denotes sublattice site $A/B$ on layer $i$. $v_i = \sqrt{3}a_0\gamma_i/2$ for $i = 0,3,4$, where $a_0 = 0.246$ nm is the lattice constant of graphene. $\gamma_i$ ($i = 0,1,...,4$) are hopping integrals between sites. $2\Delta_1$ is the potential difference between the two outermost layers and is approximately proportional to the applied displacement field. $\Delta_2$ is the potential difference between the middle layer and the average potential of the two outer layers. The dispersion relation (Fig. 1c and Fig. S2) can be obtained by diagonalizing the Hamiltonian (1). $\Delta_1$ and $\Delta_2$ depend on the experimental setup. The values of $\gamma_i$ ($i = 0,1,...,4$) and $\Delta_2$ are adopted from[2]. Suppose the energy eigenvalues of eq. (1) are $\varepsilon_i(\mathbf{k})$ ($i = 1,2,...,6$). Given $\Delta_1$ and $\Delta_2$, the DOS can be calculated by:

$$G(\varepsilon) = \int_{|k_x|,|k_y|<\Lambda} \frac{d^2 k}{2\pi^2 a_0^2} \sum_{\varepsilon_i(\mathbf{k})<\mu} \delta[\varepsilon - \varepsilon_i(\mathbf{k})], \quad (2)$$

where $\mu$ is the chemical potential and $\Lambda$ is some high-momentum cutoff constant for the low-energy effective model. In our experimental setup, one can calculate the dielectric displacement $D$ from gate voltages. Given $D$, the value of $\Delta_1$ can be calculated self-consistently[87]. The screened potential difference between outer graphene layers, $2\Delta_1$, satisfies:

$$2\Delta_1 = \frac{2Dd}{\varepsilon_r} + \frac{e^2 d}{\varepsilon_r \varepsilon_0}(n_1 - n_3), \quad (3)$$

where $d = 0.335$ nm is the distance between graphene layers, $\varepsilon_r$ is the relative permittivity between the trilayer interlayer spaces (without the screening effect of π-band electrons of the trilayer graphene), and $n_1$ and $n_3$ are the electron densities on layers 1 and 3, respectively.

Meanwhile, the values of $n_1$ and $n_3$ can be obtained by integrating the LDOS of the system described by eq. (1):

$$n_j = \int_{\varepsilon<\mu} d\varepsilon \, g_j(\varepsilon), \, j = 1,2,3. \quad (4)$$

$$g_j(\varepsilon) = \int_{|k_x|,|k_y|<\Lambda} \frac{d^2 k}{2\pi^2 a_0^2} \sum_{\varepsilon_i(\mathbf{k})<\mu} \delta[\varepsilon - \varepsilon_i(\mathbf{k})] \left(|\psi_{i,Aj}|^2 + |\psi_{i,Bj}|^2\right), \quad (5)$$

where $\psi_{i,Aj}$ and $\psi_{i,Bj}$ are the projections of the energy eigenstates corresponding to $\varepsilon_i(\mathbf{k})$ onto the $A$ and $B$ sublattices of layer $j$. We use the initial guess $\Delta_1 = \frac{Dd}{\varepsilon_r}$ and self-consistently iterate over equations (4) and (3) until convergence to obtain the relation between $D$ and $\Delta_1$.

The joint density of states (JDOS) is calculated from:

$$J(\varepsilon, \mathbf{q}) = \int \frac{d^2k}{2\pi^2 a_0^2} \sum_{\varepsilon_i(\mathbf{k}),\, \varepsilon_j(\mathbf{k+q})<\mu} \delta[\varepsilon - \varepsilon_i(\mathbf{k})]\delta[\varepsilon - \varepsilon_j(\mathbf{k})]. \tag{6}$$

**Determine the carrier density and displacement fields**

The capacitance per unit area between sample and graphite back gate $C_{\mathrm{bg}}$, can be determined accurately using quantum oscillations. In our measurement, well-defined LLs were obtained in the ABA domain (Supplementary Note 3 and Fig. S3). The degeneracy of each LL is $\frac{\Phi}{\Phi_0}$, where $\Phi_0 = \frac{h}{e}$ is the magnetic flux quantum. The capacitance of the device is extracted from the following formula:

$$C_{\mathrm{bg}} \Delta V_{\mathrm{bg}} = N \cdot \frac{4Be^2}{h}, \tag{7}$$

where $\Delta V_{\mathrm{bg}}$ is the back gate voltage change. $N = 3$ is the number of LLs accounted within the back gate range $\Delta V_{\mathrm{bg}}$, $B$ is the applied magnetic field, $e$ is the electron charge, and $h$ is Planck's constant. The calculated capacitance $C_{\mathrm{bg}}$ is approximately $2.58 \times 10^{11}$ e cm$^{-2}$V$^{-1}$ for this device. From the obtained $C_{\mathrm{bg}}$, we can determine the average perpendicular dielectric constant $\varepsilon_{\mathrm{hBN}} = 3.5$ using the expression $C_{\mathrm{bg}} d / \varepsilon_0$, where $d$ is the thickness of the dielectric layer (hBN), which is 75 nm from the AFM measurements. This capacitance corresponds to the parallel plate capacitor formed between the graphite and the sample and remains uniform across the entire sample. Therefore, the carrier density and displacement fields as a function of back gate for the sample can be determined using the capacitance obtained from quantum oscillations.

Experimentally, the relation between $n_e$, $D$ and $V_{\mathrm{bg}}$ shown in Fig. 1e can be written as follows[59–61]:

$$n_e = \frac{C_{\mathrm{bg}} V_{\mathrm{bg}}}{e} + n_0, \tag{8}$$

$$D = \frac{C_{\mathrm{bg}} V_{\mathrm{bg}}}{2\varepsilon_0} + D_0, \tag{9}$$

where $\varepsilon_0$ is the vacuum permittivity. The initial doping $n_0$ is experimentally determined from the difference between $V_{\mathrm{bg}} = 0$ and $V_{\mathrm{bg}}$ at CNP ($V_{\mathrm{bg}}(n_e = 0)$). We identify $V_{\mathrm{bg}}(n_e = 0)$ when there is a sharp slope change in the $dI/dV$ ($n_e, D$) map (Fig. 1e). This gate voltage is primarily determined by the sample geometry but may also vary slightly across different regions due to local potential or alignment with the hBN substrate. Similarly, $V_{\mathrm{bg}}$ corresponding to the zero-displacement field $V_{\mathrm{bg}}(D = 0)$ is defined by the gate voltage where the energy spacing between the VHSs of conduction and valence band is minimum, as illustrated in Fig. S3. From this we obtain the built-in displacement field $D_0 \approx 0$, $V_{\mathrm{bg}}(D = 0) \approx 0$. Parameters $n_0$ and $D_0$ for different regions are adjusted in accordance with $V_{\mathrm{bg}}(n_e = 0)$ and $V_{\mathrm{bg}}(D = 0)$ derived from the $dI/dV$ ($n_e, D$) map. For RTG region discussed in the main text, $V_{\mathrm{bg}}(n_e = 0) = -2.8$ V, indicating that the valence band pinning onsets when the gate voltage is below -2.8 V.

We then incorporate $n_0$ and $D_0$ into the theoretical framework. Using a self-consistent simulation previously described, we apply a displacement field offset of -0.065 V/nm, derived from the experimental value of $n_0$. By setting $\varepsilon_{\mathrm{hBN}} = 3.5$, as obtained from capacitance estimations, we obtain the single-particle DOS for the top layer graphene, $g(\mathrm{top})$ shown in Fig. 1f. Notably, both the layer polarization and the ranges of $n_e$ and $D$ in the simulation align well with the $dI/dV$ ($n_e, D$) map (Fig. 1e). We use the same capacitance to determine the carrier density and displacement fields $n_e$ and $D$ for low temperature measurements and parameters in all the other regions.

**Extraction of *q* vectors from QPI**

For QPI analysis near the scatterer in Figs. 2c and d, the d$I$/d$V$ map was normalized by integrated d$I$/d$V$ from $V_{\text{bias}}$ = 0 to the setpoint $V_{\text{set}}$, which can improve the signal to noise ratio in the FT analysis. We then cropped the map in the region of interest and fine-tuned the FT window to get a clear signal in FT. During this process, we ensure that the observed features remain robust against variations in the window sizes and positions. The two modes in the valence band for $V_{\text{bg}}$ = -9.5 V are consistent across different windows. For Fig. 2e (left), we normalize the raw d$I$/d$V$ map by $I/V$, which leads to clear emergence of the two modes with low noise. This allows for easier comparison with the theoretical curves. For the phase of FT analysis shown in Fig. 2e (right), we identify that abrupt $\pi$ phase jumps occur at band edges of overlapping pockets, the extent of these phase jumps as a function of energy reflects the shape of these pockets. The two $\pi$ phase jumps signify the split bands that we identify as the IVC bands. To calculate the *q-E* dispersion in Fig. 2f (solid lines), we use the SCHF bands at high hole density ($n_e = $ -1.5 × 10$^{12}$ cm$^{-2}$)[13]. For elastic scattering processes, we identify the equal-energy states of momenta **k** and **k**′ for the SCHF bands. Then we calculate the momentum transfer wavevectors **q** = **k** - **k**′ for different energies thereby reconstructing *q-E* dispersion (Fig. 2f) from *k-E* dispersion (Fig. 2g).

**Self-consistent Hartree-Fock calculations**

To understand the order instability in the system, we performed unconstrained momentum-resolved SCHF calculations, allowing for incommensurate IVC order. The model Hamiltonian is given by:

$$H = \sum_{\mathbf{k}} \psi_{\mathbf{k}}^{\dagger} H_0(\mathbf{k}) \psi_{\mathbf{k}} + H_C \qquad (10)$$

where $\psi_{\mathbf{k}}$ is a vector in sublattice space of annihilation operators of electron with momentum **k**, $H_0$ is the single particle Hamiltonian defined above, and $H_C$ accounts for gate-screened Coulomb interaction. The mean-field analysis captures a plethora of phases, both diagonal in valley isospin and with off-diagonal (IVC) order both at zero momentum (commensurate IVC) and at finite momentum (incommensurate IVC). The calculation was performed using the method described in detail in ref. 13. We extract DOS in the mean-field ground state by replacing the bare $\varepsilon_i(\mathbf{k})$ and $\psi_i(\mathbf{k})$ in Eq. (5) with their values evaluated at the SCHF ground state. We calculate the DOS along a line in the phase-space connecting the points $n_e$ = -0.2 × 10$^{12}$ cm$^{-2}$, $\Delta_1$ = 15 meV and $n_e$ = -2.2 × 10$^{12}$ cm$^{-2}$, $\Delta_1$ = 35 meV, and the results are plotted in Fig. 3.

**Data availability**

The data that support the findings of this study are available from the corresponding authors on request.

**Code availability**

The calculations codes used in this study are available from the corresponding authors on request.

**Acknowledgements**

The author would like to acknowledge discussions with M. Zaletel, W. Zhi, and we thank T. Xiao and L. Joshi for helping with the electrode preparation. This work was primarily supported by the BSF (Grant Numbers 2022272). Work at UCSB was primarily supported by the National Science Foundation under award DMR-2226850, with additional support provided by the Gordon and Betty Moore Foundation under award GBMF9471. The work also made use of shared equipment sponsored by the National Science Foundation through Enabling Quantum Leap: Convergent Accelerated Discovery Foundries for Quantum Materials Science, Engineering and Information (Q-AMASE-i) award number DMR-1906325. K.W. and T.T. acknowledge support from the JSPS KAKENHI (Grant Numbers 21H05233 and 23H02052) and World Premier International Research Center Initiative (WPI), MEXT, Japan.

**Author contributions**

N.A, H.B, A.F.Y. conceived of and directed the project. Y.L. developed the setup with the help of H.Z and B.B. Y.L. and G.A. performed the STM measurements with the help of Y.C. Y.C. and H.S. fabricated the trilayer graphene devices supervised by A.F.Y.. Y.L., G.A, H.B. and N.A. analyzed the data. Y.W. performed the tight-binding calculation supervised by B.Y.. Y.V., J.X. formulated the theory and performed the Hartree-Fock calculation supervised by E.B.. K.W. and T.T. provided the hBN crystals. Y.L., G.A., Y.V., E.B., H.B. and N.A. wrote the paper, with input from all the co-authors.

**Competing financial interests**

The authors declare no competing interests.